\input harvmac
\input epsf


\def\figin{\epsfcheck\figin}\def\figins{\epsfcheck\figins}
\def\epsfcheck{\ifx\epsfbox\UnDeFiNeD
\message{(NO epsf.tex, FIGURES WILL BE IGNORED)}
\gdef\figin##1{\vskip2in}\gdef\figins##1{\hskip.5in}
\else\message{(FIGURES WILL BE INCLUDED)}%
\gdef\figin##1{##1}\gdef\figins##1{##1}\fi}
\def\DefWarn#1{}
\def\figinsert{\goodbreak\topinsert}
\def\ifig#1#2#3#4{\DefWarn#1\xdef#1{fig.~\the\figno}
\writedef{#1\leftbracket fig.\noexpand~\the\figno}%
\figinsert\figin{\centerline{\epsfxsize=#3mm \epsfbox{#2}}}
\bigskip\medskip\centerline{\vbox{\baselineskip12pt
\advance\hsize by -1truein\noindent\footnotefont{\sl Fig.~\the\figno:}\sl\ #4}}
\bigskip\endinsert\noindent\global\advance\figno by1}

\def\b{\beta}

\def\G{\Gamma}

\def\eps{\epsilon}

\def\th{\theta}

\def\k{\kappa}
\def\l{\lambda}

\def\t{\tau}

\def\tu{{\tilde u}}
\def\th{{\tilde h}}

\def\d{\partial}

\def\hf{{1\over 2}}

\def\ai{{\rm Ai}}

\def\({\bigl(}
\def\){\bigr)}
\def\<{\langle\,}
\def\>{\,\rangle}
\def\]{\right]}
\def\[{\left[}

\lref\dvm{
R.~Dijkgraaf and C.~Vafa,
``Matrix models, topological strings, and supersymmetric gauge theories,''
Nucl.\ Phys.\ B {\bf 644}, 3 (2002)
[arXiv:hep-th/0206255].}

\lref\dvg{
R.~Dijkgraaf and C.~Vafa,
``On geometry and matrix models,''
Nucl.\ Phys.\ B {\bf 644}, 21 (2002)
[arXiv:hep-th/0207106].}

\lref\dvp{
R.~Dijkgraaf and C.~Vafa,
``A perturbative window into non-perturbative physics,''
arXiv:hep-th/0208048.}

\lref\inth{
M.~Aganagic, R.~Dijkgraaf, A.~Klemm, M.~Marino and C.~Vafa,
``Topological strings and integrable hierarchies,''
arXiv:hep-th/0312085.}

\lref\ks{M.~Bershadsky, S.~Cecotti, H.~Ooguri and C.~Vafa,
``Kodaira-Spencer theory of gravity and exact results for quantum string
amplitudes,''
Commun.\ Math.\ Phys.\  {\bf 165}, 311 (1994)
arXiv:hep-th/9309140.}

\lref\bz{
E.~Brezin and A.~Zee,
``Universality of the correlations between eigenvalues of large random
matrices,''
Nucl.\ Phys.\ B {\bf 402}, 613 (1993).}

\lref\kostov{
I.~K.~Kostov,
``Conformal field theory techniques in random matrix models,''
arXiv:hep-th/9907060}.

\lref\ecryst{
A.~Okounkov, N.~Reshetikhin and C.~Vafa,
``Quantum Calabi-Yau and classical crystals,''
arXiv:hep-th/0309208.}

\lref\cryst{A.~Iqbal, N.~Nekrasov, A.~Okounkov and C.~Vafa,
``Quantum foam and topological strings,''
arXiv:hep-th/0312022.}

\lref\nat{
N.~Saulina and C.~Vafa,
``D-branes as defects in the Calabi-Yau crystal,''
arXiv:hep-th/0404246.}

\lref\okounkov{A.~Okounkov,
``Symmetric functions and random partitions,''
arXiv:math.CO/0309074}

\lref\cs{M.~Marino,
``Chern-Simons theory, matrix integrals, and perturbative three-manifold
invariants,''
arXiv:hep-th/0207096.}

\lref\mcs{
M.~Aganagic, A.~Klemm, M.~Marino and C.~Vafa,
``Matrix model as a mirror of Chern-Simons theory,''
JHEP {\bf 0402}, 010 (2004)
[arXiv:hep-th/0211098].}

\lref\eyn{
B.~Eynard,
``Eigenvalue distribution of large random matrices,
from one matrix to several coupled matrices,''
arXiv:cond-mat/9707005.}

\lref\inff{A.~Okounkov,
``Infinite wedge and random partitions,''
arXiv:math.RT/9907127.}

\lref\schur{A.~Okounkov and N.~Reshetikhin,
``Correlation function of Schur process with application to local
geometry of a random 3-dimensional Young diagram,'' 
arXiv: math.CO/0107056.}

\lref\bh{
H.~Ooguri, A.~Strominger and C.~Vafa,
arXiv:hep-th/0405146.}

\lref\tw{
C.~A.~Tracy and H.~Widom,
``Differential Equations of Dyson Processes,''
arXiv: math.PR/0309082.}

\lref\sebas{
S.~de Haro and M.~Tierz,
``Brownian motion, Chern-Simons theory, and 2d Yang-Mills,''
arXiv:hep-th/0406093.}


\Title
 {\vbox{
\hbox{hep-th/0406247}
\hbox{ITFA-2004-22}}}
{\vbox{
\centerline{Universal Correlators from Geometry}
}}
\bigskip
\centerline{Robbert Dijkgraaf$^{1,2}$, Annamaria Sinkovics$^1$,
and Mine Tem\"urhan$^1$}
\vskip8mm
\centerline{\sl $^1$Institute for Theoretical Physics}
\centerline{\sl $^2$Korteweg-de Vries Institute for Mathematics}
\centerline{\sl University of Amsterdam}
\centerline{\sl Valckenierstraat 65}
\centerline{\sl 1018 XE Amsterdam, The Netherlands }
\bigskip
\bigskip
\vskip .1in\centerline{\bf Abstract}

Matrix model correlators show universal behaviour at short
distances. We provide a derivation for these universal correlators by
inserting probe branes in the underlying effective geometry.  We
generalize these results to study correlators of branes and their
universal behaviour in the Calabi-Yau crystals, where we find a role
for a generalized brane insertion.

\smallskip

\vfill

\Date{June, 2004}

\newsec{Introduction}

Topological strings on Calabi-Yau manifolds provide diverse
connections between string theory, geometry and random matrix
models. In particular it has been shown that B-model topological
strings on certain non-compact Calabi-Yau backgrounds can be described
in terms of matrix models \dvm, \dvg, \dvp, where the geometry emerges
from the planar limit of the random matrix integral. Here the matrix
model appears as the open strings living on compact branes.

The geometry of these B-models can be effectively analyzed by making
use of non-compact B-branes \inth. Since these branes have infinite
world-volume they are infinitely heavy and should be considered as
non-dynamical external probes. More specifically, we are considering
non-compact Calabi-Yau manifolds that are given as the hypersurface
\eqn\surf{
z w - H(y,x) =0, } 
where $z, w,y ,x \in \bf{C}$. The non-compact branes are then
parametrized by a fixed point in the $(x,y)$ plane that lies on the
curve 
$$ 
{\cal C}:\ H(x,y)=0, 
$$ 
and they extend in the coordinate $z$
or $w$.

The B-model describes the complex structure deformations of the complex
curve ${\cal C}$ or, equivalently, of the ``Hamiltonian'' $H(x,y)$.
The variations of the complex structure at infinity can be introduced by 
a chiral boson, $\phi(x)$. 
In a local coordinate patch this chiral boson is defined by
$$
y(x) = \d \phi(x),
$$
and it describes the variation of the curve through its parametrization $y(x)$.

The target space theory of the B-model on this geometry is the
Kodaira-Spencer theory \ks\ where $\d\phi$ is the dimensional reduction of the 
KS field $A$. 
The B-branes can be thought as defects for the KS field $\phi(x)$.
The operator $\psi(x)$ that creates or annihilates a brane turns out
to be a free fermion field and is related to the 
chiral boson via the familiar bosonization formula
$$ 
\psi(x) = e^{i \phi(x) / g_s}.
$$
Similarly the field $\psi^*(x) = e^{-i\phi/g_s}$ creates/annihilates an anti
D-brane.

In terms of the matrix model, the chiral boson is the collective field
of the eigenvalues of the matrix.
The fermions are basically free fermions, but transform between the
different patches as wavefunctions with generalized Fourier transformation.
We will make use of this fact by choosing good coordinates, and later
Fourier transforming back to the original coordinates.

We will use the fermionic formulation to study certain correlators 
in the matrix models. In random matrix models, it has been understood
for a long time that eigenvalue correlators have an interesting behaviour
at short scales, called ``universality'' \bz. In particular, the joint 
probablity of $n$ eigenvalues is given by a determinant of a single  
kernel 
$$
 \rho(\l_1, \ldots \l_n) = \det_{n \times n} K(\l_i, \l_j). 
$$
In the limit $N$ large, while keeping the rescaled distances
$N(\l_i - \l_j)$ fixed, the kernel $K$ takes the form
\eqn\sinker{
K(\l_i, \l_j) \sim {\sin{N \pi \rho(\bar{\l}) (\l_i - \l_j)} \over 
N \pi (\l_i -\l_j)}, }
where $\rho({\bar \l})$ is the density of eigenvalues at the mean ${\bar \l}(\l_i + \l_j) /2$.
The formula is called universal because it does not depend explicitly on
the form of the matrix model potential, thus it has the same form for
any potential. It has only a functional dependence on the potential 
through the relatively uninteresting scaling factor of the mean
density.

There are many ways to derive this kernel in random matrix models,
the usual one relies on introducing orthogonal polynomials. With the
use of brane insertions, we provide a simple derivation by writing the
kernel in terms of the correlator of free fermions as
\eqn\mainker{
K(x_1, x_2) = \langle \psi(x_1) \psi^{*}(x_2) \rangle \sim 
{{e^{i (\phi(x_1) - \phi(x_2)) / g_s} \over x_1 - x_2} 
}, }
where the second equality is by bosonization. Here $\phi(x)$ is
the chiral boson which is related to the geometry associated to the 
matrix model. In this formula, the use of good coordinates, which
are single valued, is an essential point. That is, it only takes this
simple form if the coordinate $x$ parametrizes a unique point on the 
curve ${\cal C}$. It is only in these coordinates
the branes, or free fermions, can be inserted at a definite position.

As we will explain, to find the kernel in the usual double valued
coordinates, we can use Fourier transformation. In fact, we are just
applying standard techniques of the semiclassical WKB
approximation. The kernel will then be given as a weighted sum of
contributions, coming from the ``images'' of the brane in the
multivalued coordinates. In the multicover coordinates, the position
of the brane cannot be fixed unambigously, and in fact, the brane
insertion is best defined via the Fourier transformation.  A version
of this idea first appeared in \kostov\ using methods of conformal
field theory. Here we make it more explicit and reformulate it in the
recent language of brane insertions.

The universal correlator is an interesting quantity to consider, since
it probes the geometry at short distances. We can investigate it with the idea
of inserting probe branes of the geometry. In fact, brane probes can
also be used in the more general context of the Calabi-Yau crystal \ecryst,
\cryst, \nat.

We study certain correlators in the crystal which are the 3 dimensional 
analogues of the eigenvalue correlators of matrix models. We find that
such correlators can be described by fermion insertions in the geometry.
These fermions are ''generalized`` in the sense they depend on two parameters.
They are the usual one dimensional chiral fermions, but inserted at an
arbitrary slice of the crystal, giving an additional parameter.
Thus the fermions are effectively two dimensional objects probing the
3 dimensional structure of the crystal.

The generalized fermions do not correspond to the Lagrangian brane probes
in \nat. They are different objects probing the interior
of the crystal. Finding the precise brane interpretation is an interesting
open problem.

It is natural to ask whether the free fermion correlators of the crystal 
also show a similar universal behaviour when scaling parameters appropriately.
In fact such correlators have been computed in the mathematics literature
of random partitions \okounkov, and a similar structure of universal 
correlators is found. 

When taking a two dimensional slice of the crystal, in the scaling region
we find the same sine kernel as in the case of random matrices. 
It is well-known in the mathematics of partitions that one can define
a suitable probability measure which is the discretized version of
the measure for Gaussian matrix models. The three dimensional analogue of
the sine kernel is a more complicated object, written in terms of an
incomplete beta-function. We conjecture this beta-function is the 
universal scaling limit of a certain 2-matrix model with a unitary
measure, which appears in the context of Chern-Simons theory \cs, and
the topological vertex \mcs.

\newsec{Branes and the WKB approximation}

To understand the relation between eigenvalues in matrix models and
brane insertions in topological strings, let us first review some
familiar facts about semi-classical quantization.  We consider $(x,y)$
as coordinates on a two-dimensional phase space. The curve ${\cal C}$
given by $H(x,y)=0$ can be viewed as the orbit that corresponds to the
classical ground state of the system described by the Hamiltonian
$H$. To such a one-dimensional curve one can associate the (global)
action variable $\mu$ given by the contour integral
$$
\mu = {1\over 2\pi} \oint_{\cal C} y dx. 
$$
The Bohr-Sommerfeld rule quantizes this as $\mu = (N+ {1\over 2})
g_s$, where we identify here and subsequently $\hbar$ with the string
coupling constant $g_s$. It is useful to think of the Hamiltonian
$H(x,y)$, and therefore also the curve ${\cal C}$, as part of a family
of systems parametrized by the modulus $\mu$.

\ifig\orbit{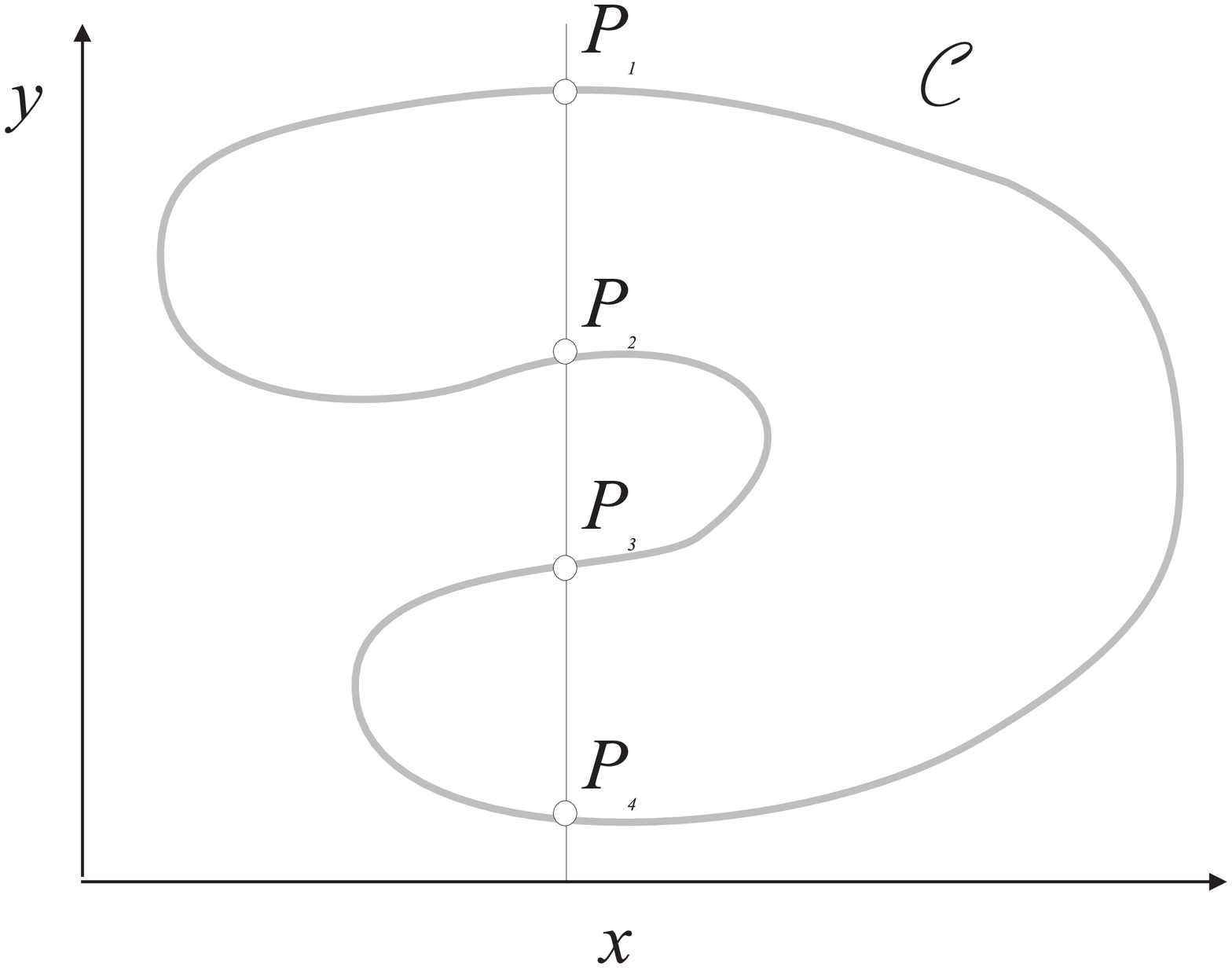}{60}{
The graph of the function $y(x)$ is given by the orbit ${\cal C}$
determined by $H(x,y)=0$. In general it has locally several branches:
to a single value of $x$ correspond the geometric points
$P_1,P_2,\ldots$.}

In general, if we want to use $x$ as a coordinate and solve for $y$ as
a function of $x$ on the curve $H(x,y)=0$, there can be several
branches $y_I(x)$. This is because the line of constant $x$ will
intersect the curve ${\cal C}$ in different points $P_I$, see \orbit.

All these points at these different sheets will contribute to the
semi-classical wavefunction $\psi(x)$ associated to the orbit $\cal
C$. The phase of this wavefunction is given by the local action
$$
\phi_I(x) = \int_{x_0}^x y_I(x') dx'
$$
with $x_0$ some arbitrary reference point. The full WKB expression for
the wave function reads with some overall normalization $c$
$$
\psi(x) = c \cdot \sum_I 
\left({\partial^2 \phi_I \over \partial x \partial \mu}\right)^{1/2}
\exp\left[i { \phi_I(x) \over g_s }\right]
$$
Here all the intersection points $P_I$ contribute. For example, for
the harmonic oscillator with 
$$
H(x,y) = x^2 + y^2 - 4 \mu
$$
we have two such points. Indeed the wavefunction is well-known to be
given in the WKB approximation by (for the ``allowed zone'')
$$
\psi(x)  \sim (4\mu - x^2)^{-{1\over 4}} 
\cos \left(\int^x \!\!\sqrt{4\mu -x'^2} dx' - {\pi\over 4}\right).
$$

In the context of matrix models for the topological B-model, the
geometric points $P_I$ on the curve ${\cal C}$ correspond to
non-compact D-branes on the Calabi-Yau space. The variable $x$ appears
naturally in the matrix model as the eigenvalue, but will not be a
good coordinate, since the spectral curve ${\cal C}$ will always be a
multiple cover over the $x$-plane (a double cover for a single matrix
integral). The natural fermion operators $\psi(x)$, that appear in the
matrix model and correspond to creating a single eigenvalue, are
therefore not represented by a single geometric brane. Instead they
are a superposition of branes inserted at the inverse images $P_I$ of
$x$, just as we have in the formula for the WKB wavefunction.

In particular for the local geometry with two sheets, where we forget
about all the $x$-dependence and therefore can write the local
geometry as
$$
y^2 = p^2,
$$
we have branes inserted at the points $P_1$ and $P_2$ given by
$y_{1,2} = \pm p$. The corresponding action functions are
$\phi_1(x) = p x$ and $\phi_2(x) = -p x$. The
wavefunction is naturally given as the sum
$$ 
\psi(x) \sim p^{-1/2} \left[
\exp i\left({p x \over g_s} + {\pi\over 4}\right)
+ \exp i\left(-{p x \over g_s} - {\pi\over 4}\right)
\right].
$$ 
That is, we have in terms of brane insertions the natural
identification
$$ 
\psi(x) = e^{i\pi\over 4} \psi(P_1) + e^{-{i\pi\over 4}} \psi(P_2).
$$ 
Similarly the anti-brane is inserted by
$$ 
\psi^*(x) = e^{i\pi\over 4} \psi^*(P_1) + e^{-{i\pi\over 4}} \psi^*(P_2).
$$ 
(Note that this is not the complex conjugated expression.)

With this translation we can easily compute multi-fermion correlation
function. Here we should remind us that we will have only contraction
of operators that create or annihilate branes on the same sheet.  For
example, we have the two-point function
$$
\left\langle \psi(x) \psi^*(x') \right\rangle = 
e^{i\pi\over 2} \left\langle \psi(P_1) \psi^*(P'_1) \right\rangle - 
e^{-{i\pi\over 2}} \left\langle
\psi(P_2) \psi^*(P'_2) \right\rangle 
$$
This can be evaluated in the limit $x \to x'$ as
$$
\left\langle \psi(x) \psi^*(x') \right\rangle \sim  
{\sin\left({\sqrt\mu(x - x')/ g_s}\right)
\over  \sqrt\mu (x-x')}
$$
which gives the famous sine-kernel.

\ifig\turning{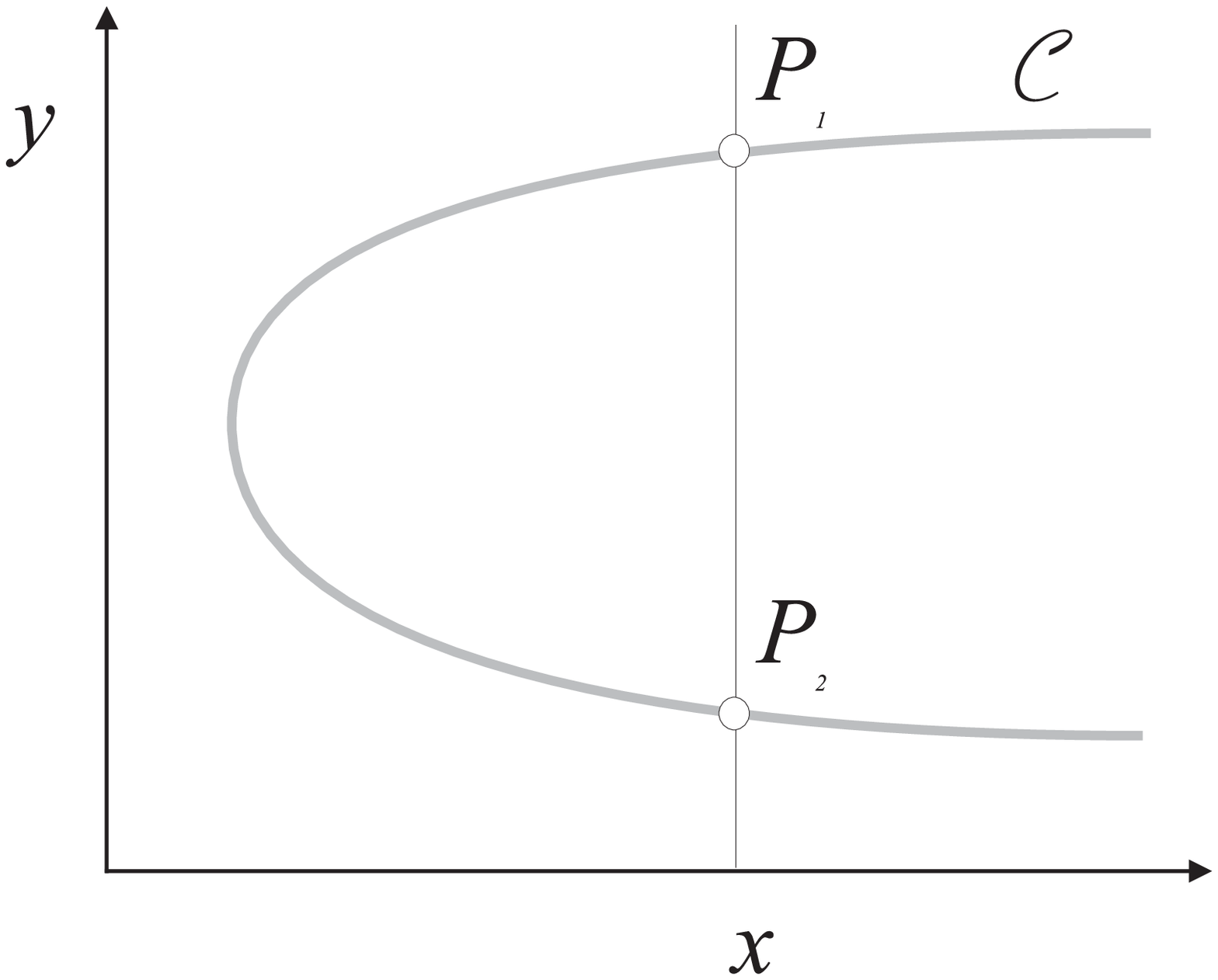}{50}{
Close to a turning point, given by $y^2=x$, the good coordinate is $y$
not $x$.}

This description becomes problematic at caustics or turning points
where the curve ${\cal C}$ is perpendicular to the $x$-direction and
two of the branches coalesce as in \turning. The right description at
these points is to use the coordinate $y$ instead. The new
wavefunction is then determined by a Fourier transform
$$
\psi(y) = \int dx \ e^{ixy/g_s} \psi(x).
$$
As we will see later this naturally leads to the Airy kernel.

\newsec{Universal correlator in random matrix models}

We now turn to a detailed investigation of the universal correlators
in matrix models using the geometric picture of brane insertions just
described.
Consider an $N\times N$ matrix model with partition function

\eqn\MM{
Z = \int d \Phi \ e^{-{1 \over g_s} {\rm Tr} W(\Phi)}}
and let us define the density of $n$ of its eigenvalues as 
\eqn\dens{
\rho(\l_1, \ldots \l_n) = {(N-n)! \over N!} \Big\langle \prod_{k=1}^n
{\rm Tr}\ \delta (\l_k - \Phi) \Big\rangle.}
For the random matrix model this describes the probability of measuring
$n$ eigenvalues at the same time. More precisely, the interesting part 
of this correlation function is its connected part.
In the large $N$ limit the multicorrelator factorizes into one point functions,
and therefore only its connected component contains new information.

Introducing orthogonal polynomials
$$ \int d \l \ e^{-{W(\l) \over g_s}} P_n (\l) P_m(\l) = \delta_{nm} $$
the joint eigenvalue distribution can be expressed \bz, \eyn\ as 
a determinant a single two-point kernel
\eqn\kdet{\rho_{n}(\l_1, \ldots \l_n) = {N^n (N-n)! \over N!} \det_{n \times n} 
K(\l_i, \l_j).}
In terms of the wavefunctions
$$ \Psi_n(\l_i) = P_n(\l_i) e^{-{W(\l) \over 2 g_s}}  $$
this kernel can be written as
\eqn\kerp{
K(\l_i, \l_j) = {1 \over N} \sum_{n=0}^{N-1} \Psi_n(\l_i) \Psi_n(\l_j)
= { \Psi_N(\l_i) \Psi_{N-1}(\l_j) - \Psi_{N-1}(\l_i)
\Psi_N(\l_j)  \over \l_i - \l_j}.}
Here we dropped the overall normalization factors, and at the second
equality used the Darboux-Christoffel formula for the 
orthogonal polynomials.
This formula can also be interpreted as a Slater determinant for second 
quantized fermions
$$
K(\l_i, \l_j) = \langle \Psi^{\dagger}(\l_i) \Psi(\l_j) \rangle.
$$ 

The eigenvalues of matrix model are usually located along a cut on the
real axis. When scaling close to a fixed point in the middle of the cut,
this kernel satisfies universal properties
\eqn\uc{K(\l_i, \l_j) \sim {\sin{\[N \pi (\l_i - \l_j) 
\rho(\bar{\l})\]} \over N \pi (\l_i - \l_j) }} 
in the region where $(\l_i - \l_j) \sim O(1 / N)$. Here
$\bar{\l} = (\l_i + \l_j)$/2, and $\rho({\bar {\l}})$ is the
single eigenvalue density. 

When scaling close to the endpoint of the cut, the kernel has a different
expression in terms of Airy functions
\eqn\ucAi{K(\l_i, \l_j) \sim {\ai(\l_i) \ai'(\l_j) - \ai(\l_j) \ai'(\l_i)
\over \l_i - \l_j},
}
where the $\l's$ are now suitably scaled variables close to the endpoint.

Since the form of the kernel does not depend on the specific form of 
the potential, we can restrict attention to the case of Gaussian potential
$V(x) = x^2/2$. 
For the Gaussian potential, the sine and the Airy kernel can be 
derived by using the direct asymptotic expansion
of the wavefunctions, which in this case are Hermite polynomials. 
We include this standard computation in the Appendix.

\newsec{Universal correlator from geometry}

In the recent context of topological string theory, matrix models
are directly related to the geometry of the CY. 
In the limit $N \rightarrow \infty$, 
$g_s \rightarrow 0$, $\mu=g_s N$ fixed, the CY geometry associated to
the matrix model is
$$zw + y(x)^2 + (W'(x))^2 - f(x) = 0.$$
Here the quantum deformation $f(x)$ is a polynomial computed from the 
potential as
$$
f(x) = {4 \mu \over N} \sum_{i} {W'(x) - W'(\l_i) \over x - \l_i}, 
$$
and 
$$y(x) = \d \phi(x)$$
is the chiral boson from the collective field
$$ \phi(x) = 2 g_s {\rm Tr} \log(x- \Phi) - W'(x).$$

The Riemann surface associated with the Gaussian matrix model in this
way is given by
\eqn\geom{
H(y,x) = y^2 + x^2 - 4 \mu = 0.
}
As discussed in \inth, the geometry can be probed by inserting non-compact
branes. Taking an asymptotic patch in the geometry, and considering 
the worldvolume action of the noncompact brane one finds $x$ and $y$
are a symplectic pair of variables with canonical commutation relations
$[x, y] = i g_s$. In terms of operators, 
$$
y= -i g_s {\d \over \d x}.
$$ 
Inserting a non-compact brane in this geometry corresponds to adding
a fermion $\psi(x) = e^{i \phi(x) / g_s}$ with wavefunction
\eqn\wavef{
\Psi(x) = \langle \psi(x) \rangle = e^{i \phi_{\rm cl} (x) / g_s}.
}
This wavefunction satisfies the Schr\"odinger equation
\eqn\schr{
H\left(-i g_s {\d \over \d x},  x \right) \Psi(x) = 0.
}
For the Gaussian matrix model these wavefunctions are Hermite polynomials.
In fact in the Gaussian case these are the same wavefunctions as the 
orthogonal polynomial wavefunctions of the previous section.
(In the general case the orthogonal polynomial wavefunctions do not
satisfy the Schr\"odinger equation, due to more complicated normal
ordering issues.) 

In this case, we can simply derive the kernel by finding the wavefunctions
from the geometry, and inserting in the Darboux-Christoffel formula.
Let us first show this for the Airy kernel. 
When close to the branch points, the curve is simply
\eqn\curveb{
y^2 + x =0,  
}
where we rescaled $x$ for convenience, and put the branch point at
$x=0$.
We can insert a brane in a definite position in the $y$-coordinate which
is single valued.
In the $y$ coordinate the wavefunction solves
$$ 
\left( y^2 + i g_s {\d \over \d y} \right) \Psi(y) =0,
$$
so it is given by 
$$ \Psi(y) =  e^{i {y^3 \over 3 g_s}}. $$
(where we dropped the overall constant.)
Transforming back to the $x$ coordinates we get
$$\Psi(x) =  \int dy \ e^{{i x y \over g_s}} \Psi(y) = {g_s}^{1/3} 
{\rm Ai}\[ {g_s}^{-2/3} x \]. $$
In the large $N$ limit the Darboux-Christoffel formula becomes
$$
K(x_1, x_2) \sim {\Psi(x_1) \Psi'(x_2) - \Psi(x_2) \Psi'(x_1) \over
x_1 - x_2},
$$
where $\Psi(x)$ are now the wavefunctions defined from the planar geometry,
keeping $N$ large and $\mu=g_s N$ fixed.
So from the Darboux-Christoffel formula we immediately have
$$
K(x_1, x_2) \sim {{\rm Ai}(x_1) {\rm Ai'}(x_2) - {\rm Ai}(x_2) {\rm Ai'}(x_1)
\over x_1 - x_2}.
$$
where $x_{1,2}$ are rescaled coordinates measuring distances from the 
branch points.

The sine kernel can be derived in a similar way by scaling to a fixed
point in the cut away from the branch points. In this case the curve 
reduces to
$$ 
y^2 + {\bar x}^2 - 4 \mu + 2 x \bar{x} =0, 
$$
where $\bar{x}$ is the point we scale to, and $x$ is the distance to that
point. Let us work again in the $y$ coordinates. If
dropping the small x-term, the the wavefunction solves
$$ 
(y^2 - p(\bar{x})^2) \Psi(y) =0,
$$
where $p(\bar{x}) = \sqrt{4\mu - \bar{x}^2}$ is the classical momentum 
at $\bar{x}$. The wavefunction is a sum of $\delta$-functions
$$\Psi(y) = {e^{ i \pi \over 4}} 
\delta(y -p) + e^{-{i \pi \over 4}} \delta(y + p). $$
The phases are fixed using the WKB approximation discussed in section
2. Fourier transforming back
$$\Psi(x) =  e ^{{i  \pi \over 4}} e^{{i p x \over g_s}} + 
e^{-{i \pi \over 4}} e^{-{i p x \over g_s}}.$$ 

Choosing the symmetric combination and substituting back in the Slater
determinant formula we get the sine kernel. 
  
A more precise way is to keep the small $x$-term,
$$
\left(y^2 - p(\bar{x})^2  + 2 i g_s \bar{x} {\d \over \d y} \right) 
\Psi(y) =0.
$$
The corrected solution is
$$ 
\Psi(y) = \exp{\[{i \over 2 g_s {\bar x}} \left( {y^3 \over 3} - p(\bar{x})^2 y
\right) \]}. 
$$
Transforming back to $x$ the solution is again Airy function
\eqn\airyx{
\Psi(x) = (2 g_s \bar{x})^{1/3} {\rm Ai} \left[ \left({2 \bar{x} 
\over g_s^2}\right)^{1/3} \left( 
x - {p(\bar{x})^2 \over 2 \bar{x}} \right) \right]. }
Now we still have to expand in $x$, which is the small distance to the
fixed point $\bar{x}$. Doing a saddle point expansion of the Airy integral 
we pick up the contribution from the two saddle points
$$
y_{1,2} = \pm \sqrt{\bar{p}^2 - 2 x \bar{x}}.
$$
Substituting in the Darboux-Christoffel formula 
$$
K(x_1, x_2) \sim {\sin p (x_1 - x_2)  \over x_1 - x_2}
$$
we again get the sine kernel.

\newsec{Correlator from two brane insertions}

An even shorter way to derive the universal correlators is to insert
two branes in the geometry. In the single valued coordinates, where
the branes can be inserted in definite position, the kernel can be
expressed as \kostov
\eqn\kerf{
K(y_1, y_2) = \langle \psi(y_1) \psi^{*}(y_2) \rangle.
}
An intuitive derivation of this formula would express the joint 
correlators in the collective field, and then rewrite it with bosonization.
However, the $x$-coordinates here are really inconvenient, because they are
double valued, and an inconvenient averaging is involved. 
This averaging was first described in \kostov, and can be understood
as summing over the images of the brane in the multicover coordinate,
as discussed in detail in section 2.

In the single-valued $y$-coordinate, we start with the formula $\kerf$, 
which can now be interpreted
as an insertion of a brane and an antibrane in the geometry. 
The branes inserted are free fermions, consisting of a classical
wavefunction, and a quantum part
$$
\psi(y) = \Psi(y) \psi_{{\rm qu}}(y) = e^{{i \phi_{{\rm cl}}(y) \over g_s}} 
\psi_{{\rm qu}}(y).
$$ 
So the short-distance correlator in the $y$-coordinates is
$$\langle \psi(y_1) \psi^{*}(y_2) \rangle = 
{e^{{i \over g_s}(\phi(y_1) - \phi(y_2))} \over y_1 - y_2}.$$

Consider now again the geometry close to the branch point, 
$$
y^2 + x = 0.
$$
The classical wavefunction from the geometry is
$$ \Psi(y) = e^{{i \over g_s}\phi(y)} = e^{{i \over g_s}  y^3 / 3 }.$$
Transforming back to the $x$-coordinates
$$\langle \psi(x_1) \psi^{*}(x_2) \rangle = \int d y_1 d y_2 \  {\exp{ {i \over
g_s}  \left( x_1 y_1
- x_2 y_2+ y_1^3 / 3  -  
y_2^3 / 3 \right)} \over y_1 - y_2}.$$
The Fourier transform is easily evaluated by noting that
$$
e^{i x_1 y_1 - i x_2 y_2} = {-i \over x_1 - x_2} ( 
\ \d_{y_1} + \d_{y_2})\ e^{i x_1 y_1 - i x_2 y_2}
$$
Substituting and after a partial integration in rescaled coordinates we 
arrive at
$$ \langle \psi(x_1) \psi^{*}(x_2) \rangle \sim {{\rm Ai}(x_1) {\rm Ai}'(x_2)
- {\rm Ai(x_2)}{\rm Ai'(x_1)} \over x_1 - x_2}.$$

Let us expand this formula in the limit where, while close to the endpoint, 
the distance between the two points is very small, so $x_1 = \bar{x} + x$, 
$x_2 = \bar{x} -x$, where $x$ is the small distance. Doing a saddle point 
expansion of the Airy integral we pick up the contribution of the two saddle 
points $y = \pm \sqrt{\bar{x}}$.
Substituting we arrive at the sine kernel formula
$$
\langle \psi(x_1) \psi^{*}(x_2) \rangle \sim { \sin{x \sqrt{\bar{x}}} \over x}.
$$
This is consistent with the previous expression for the sine kernel, since
close to the endpoint $y^2 + x =0$ the density scales as $\rho(\bar{x}) 
\sim \sqrt{\bar{x}}$.
Doing the saddle point expansion of the Fourier transform, and picking up
the contributions from the two saddle points, we automatically
introduced an averaging in the double valued $x$-coordinates. 

We could also consider the more general geometry
$$
y^m + x =0. 
$$
In this case the wavefunctions in the y-patch are
$$
\Psi(y) =  \exp{\left({i y^{m+1} \over g_s (m+1)} \right)}.
$$
Fourier transforming in the x-patch this gives
$$
\Psi(x) =  \int dy \ \exp{ {1\over g_s}
\left(i x y + i {y^{m+1}  \over (m+1)} \right)}. 
$$
The two point correlator and its scaling limits can be similarly derived. 

\newsec{Correlators in crystal melting}

Recently a connection has been discovered between the statistical mechanics 
picture of crystal melting and A-model topological strings on Calabi-Yau manifolds.
The crystal decribes the toric base of the Calabi-Yau, with lattice spacing
of order $g_s$. The temperature of the melting crystal is $1/g_s$ \ecryst, 
\cryst. 
At very large scales the Calabi-Yau is described in terms of classical geometry.
Decreasing the distance scale to the string scales, the geometry takes the form
of a smooth limit shape. At even smaller distances a ``gravitational foam'' picture 
emerges.

The structure of the crystal can be probed by putting probe branes in
the geometry. The effect of adding a single brane probe was first studied in
\cryst, where the correction to the partition function was discussed.

Addition of multiple branes and antibranes was considered recently in
\nat. In this paper, a statistical interpretation of non-compact probe 
branes as defects in the Calabi-Yau crystal is given, and correlators for 
inserting branes are derived.
These results are in accordance to that non-compact brane probes can be 
thought of as adding free fermions with a certain geometric transformation law 
\inth, at the B-model side. 

Similarly as for the random matrices, we will investigate certain 
correlators by inserting branes in the geometry. The correlators 
we are interested in are the 3D crystal analogues of the eigenvalue
correlators of random matrices. 

\subsec{Partitions on the plane} 

Let us first consider partitions on the plane. If we draw the diagram of a
partition of $N$, and rescale it with $\sqrt{N}$, a smooth limit
curve of the diagram emerges \okounkov, \inff. 
This is the two-dimensional analogue of the limit
shape of the melting crystal. In general, the boundary is a graph of a piecewise
linear function, consisting from jumps ups and downs. Correlators in this 
context measure the probability of a given pattern of up and down jumps. 
In fact it is enough to consider the probablities to measure the set of
down jumps, since this already gives the complete information about the shape
of the partition.
When scaling $N$ large, so that the elements in the set of downs $x_i$
$$
{x_i \over \sqrt{N}} \in [-2,2]
$$
such probabilities are given by the determinant of
a single discrete sine kernel \okounkov, \inff
\eqn\sinekern{
 K_{\sin}(x_i, x_j)  =  {\sin \phi_i (x_i - x_j) \over \pi (x_i - x_j)} 
\quad  \phi_i  =  \cos^{-1} \left( {x_i \over 2 \sqrt{N}} \right)}
Thus the same sine kernel as for random matrices emerges. The important
difference however is that in this case $x_i$ are discrete, while in the random
matrix case we had a continuous distribution. Also the scaling factor $\phi$
is different from the random matrix case where we had a scaling with the density,
but such scalings can always be absorbed in the redefinition of the variables.

The endpoint formula for the random matrix correlator in terms of Airy function
also has a counterpart in random partitions.

\subsec{3D Partitions} 

We will now study similar correlators in 3D partitions.
In the 3D generalization we can build a partition from a sequence of 
diagonal slices, which satisfy the interlacing condition,
\cryst, \ecryst, \okounkov. 
 
Another way of looking at the partition is viewing it from the top, 
and projecting it to a two-dimensional tiling pattern
\foot{The projection is best understood in terms of pictures, see
for example \okounkov, \inff.}. If the partition is
located in the positive corner of $(x,y,z)$ plane, the position
of plaquettes in the tiling pattern is given in the discrete coordinates
$$
\eqalign{
t &= y-x, \cr
h &= z - {1 \over 2} (y + x).
}
$$
One can then ask for the probability of measuring a set of fixed plaquettes, at
positions $\{(t_1, h_1) \ldots (t_n, h_n)\}$.
These correlators are given as a determinant of a 3D kernel \schur
\eqn\kerfull{
\eqalign{
&K_{3D}((t_i, h_i), (t_j, h_j)) = \cr 
& = {1 \over (2 \pi i)^2} \int_{|s| = 1 \pm \eps} 
\int_{|w| = 1 \mp \eps} d s d w \ {\Phi_{3 D}(t_1, s) \Phi^{-1}_{3D}(t_2, w) 
\over s - w} {1 \over s^{h_i + {|t_i| \over 2} + {1 \over 2}} w^{-h_j 
- {|t_j| \over 2} + {1 \over 2}} }.
}
}
Here  the contour integral the top/bottom $\pm$ signs are valid when $t_1 \ge t_2$ and 
and $t_1< t_2$ respectively, $\Phi_{3D}^{-1} = 1 / \Phi_{3D}$, and 
\eqn\pphi{
\Phi_{3D}(s, t) = {\prod_{m > max(0,-t)} \ ( 1 - q^m / s) \over
\prod_{m>max(0,t)} \ (1 - q^m s) }, \quad m \in {\bf Z} + \hf. 
} 
This kernel measures the probability of finding two fixed plaquettes in
the random tiling pattern.
In fact, this two-point correlator can be thought of as a non-compact brane and
anti-brane probe insertion in the geometry. 

The A-model geometry is related by mirror symmetry to B-model on the 
mirror Calabi-Yau \inth
$$
z w - e^{-u} - e^{v} + 1  =0
$$
Inserting a single brane, the wavefunction is an eigenfunction of the
Hamiltonian
\eqn\ham{
\eqalign{
&H(u, v) \ L(u, q) \,=\,0, \cr
&H(u, v) \,=\, q^{-1/2} e^{-u} + e^{-g_s \d_{u}} -1,  \quad q = e^{-g_s}.
}
}
The one-point function which satisfies this equation is the quantum dilogarithm
\eqn\dilog{
L(u, q) = \prod_{n=0}^{\infty} \ (1 - e^{-u} q^{n + \hf}).
}
Similarly, the one point function for an antibrane is $L^{*}(u) = 1 / L(u)$.

Let us introduce the new variables
$$
\eqalign{
s &= e^{u}, \cr
w &= e^{{\tilde u}}, \cr
\th_i &= h_i - \hf + {|t_i| \over 2},\cr
\tilde{\th}_j &= h_j + \hf + {|t_j| \over 2}.
}
$$
In terms of these variables, the kernel can be rewritten as
\eqn\kerff{
K_{3D} (\th_i, \tilde{\th}_j) = 
{1 \over (2 \pi i)^2} \int du d \tu \ {1 \over e^{u} - e^{\tu}} 
e^{-u \th_i}\, e^{\tu \tilde{\th}_j}\, \Phi_{3D} (u, t)\, 
\Phi^{-1}_{3D} (\tu, t).
}
This is essentially the same form as the two-point free-fermion function of 
a brane and antibrane considered in \inth, now Fourier-transformed in the 
variables $(\th_i, \tilde{\th}_j)$. Here $\Phi_{3D}$ and $\Phi^{-1}_{3D}$ are 
the brane and antibrane one-point functions,
$$
\eqalign{
\Phi_{3D}(u, t) &= \left\{ { {\textstyle L(u,q) \over \textstyle 
L(-u + t g_s, q)} \quad t \ge 0, 
\atop
{ \textstyle L(u - t g_s, q) \over  \textstyle L(-u, q)}  \quad t < 0} 
\right. \cr
\Phi^{-1}_{3D} (\tu, t) &= {1 \over \Phi_{3D} (u, t)}.
}
$$
The formula shows that shifting the coordinate $t$ shifts us to 
another slice of the crystal, parametrized by $t$. In particular the 
wavefunction at the slice $t=0$ is

$$ \Phi_{3D}(u, 0) = {L(u,q) \over L(-u,q)}. $$

\subsec{Brane picture}

It is a natural question what is the precise brane configuration which
reproduce the 3D correlators, and how these branes are inserted in
the geometry. To understand this, we consider the picture of \nat\
where Lagrangian branes were described as defects in the crystal.

Consider now the 3D crystal constructed from a sequence of interlacing
2d partitions, ${\mu(t)}$ 
\foot{Here we follow the notation of \ecryst\ and \nat.}.
A single 2d partition is a nonincreasing
sequence of non-negative integers $\mu = \{\mu_1, \mu_2, \ldots\}$. To 
build a 3D partitions, such sequences are labeled by an integer $t$,
and must satisfy the interlacing condition
$$ 
\eqalign{
&\mu(t) < \mu(t+1)  \quad t < 0,\cr
&\mu(t+1) < \mu(t) \quad t \ge 0, 
}
$$
where $\mu(t+1) > \mu(t)$, if
$$
\mu_1(t+1) \ge \mu_1(t) \ge \mu_2(t+1) \ge \mu_2(t) \ldots
$$
A single two-dimensional slice can be described as a state in the fermionic
Fock-space. By introducing the set of down jumps
of the tableaux and its transpose
$$
\eqalign{
a_i &= \mu_i - i + \hf, \cr
b_i &= \mu_i^{T} - i + \hf,}
$$
the fermionic state describing the tableaux consisting of $d$ boxes is
given by
$$
|\mu \rangle = \prod_{i=1}^{d} \psi^{*}_{-a_i} \psi_{-b_i}.
$$
Using bosonization, we introduce
$$
q^{L_{0}} |\mu \rangle = q^{|\mu|} |\mu \rangle,
$$
where $|\mu|$ denotes the number of boxes, and the creation/annihilation
operators
$$
\Gamma_{\pm}(z) = e^{\phi_{\pm}(z)}.
$$
Here $\phi_{\pm}(z)$ are the positive and negative mode part of the chiral
boson $\phi(z)$, which is related to the complex fermion by bosonization
$$
\psi(z) = : e^\phi(z) :.
$$
In this language, introducing a D-brane corresponds to the insertion
of a fermion. Discarding the zero mode part, the D-brane operator can be
written as
$$
\Psi_{D}(z) = \Gamma_{-}^{-1}(z) \Gamma_{+} (z). 
$$
We use here and throughout the paper the standard framing $p=0$.

The partition function of the crystal with the operators can be written
as
\eqn\part{
Z(q) = \langle 0| \ \prod_{n=1}^{\infty} \Gamma_{+}(q^{n- \hf}) 
\prod_{m=1}^{\infty} \Gamma_{-}(q^{-m + \hf}) \ |0 \rangle,
}
where $q= e^{-g_s}$. With the commutation relation
\eqn\comm{
\Gamma_{+}(z)\, \Gamma_{-}(z') = (1 - z/z')^{-1} \Gamma_{-}(z')\, 
\Gamma_{+}(z),  
}
it is straightforward to show that the partition function is the 
McMahon function
$$
Z(q) = \prod_{n=1}^{\infty} (1- q^n)^{-n} = M(q).
$$

In \nat, Lagrangian D-brane insertions were considered. These branes end on
the axis, and have the geometry
$$
y= x+ a = z+ a,  \quad a > 0,
$$
where we chose the brane to end at $y=a$. 
In \nat\ it was shown that introducing such a Lagrangian brane at a position
$a = g_s (N_0 + \hf)$, corresponds to inserting a fermion D-brane
operator $\Psi_D(q^{-(N_0 + \hf)})$ at the slice $t=N_0 +1$,
$$
\eqalign{
Z_{D}(q, N_0) &= \langle 0|\  \prod_{n=1}^{\infty} \Gamma_{+}(q^{n- \hf})
\times \cr
& \prod_{m=1}^{N_0 + 1} \Gamma_{-}(q^{-m + \hf}) \Psi_D(q^{-(N_0 + \hf)})
\prod_{m=N_0 +2}^{\infty} \Gamma_{-}(q^{-m + \hf}) \ |0 \rangle.
}
$$
Commuting through the operators gives the brane one-point function
$$
\eqalign{
Z_D(q, N_0) &= M(q) \ \xi(q) \prod_{n=1}^{\infty} 
(1 - e^{-g_s (N_0 + \hf)} q^{n- \hf}) \cr
&= M(q) \xi(q) L(g_s(N_0 + \hf), q),
}
$$
where we have expressed the wavefunction in terms of the overall McMahon
function $M(q)$ and the quantum dilogarithm $L(u, q)$ we introduced 
before, and
\foot{$\xi(q)$ is a normalization factor with respect to the string answer, 
so that $Z_{{\rm crystal}} = \xi(q) Z_{{\rm string}}$ \nat.}
$$
\xi(q) = \prod_{n=1}^{\infty} {1 \over 1 - q^n}.
$$

These brane insertions depend on the single parameter $N_0$, expressing
the fact that the branes correspond to the Lagrangian geometry. 
Comparing with the brane insertions for the 3D correlators in the 
previous section, we see that our brane insertions are of more
general kind, since they depend on the two independent parameters
$t$ and $u$. In fact, they correspond to the insertion of a D-brane
operator $\Psi_D(z)$ at an arbitrary slice $t$. 

Let us insert a fermionic operator $\Psi_D(q^{-(N + \hf)})$ at the slice
$t=N_0 +1$. Note that $N$ and $N_0$ are now independent. This is given
by the operator expression
$$
\eqalign{
Z_D(q, N, N_0) &=  \langle 0| \ \prod_{n=1}^{\infty} \Gamma_{+}(q^{n- \hf})
\ \times \cr
& \prod_{m=1}^{N_0 + 1} \Gamma_{-}(q^{-m + \hf}) \Psi_D(q^{-(N + \hf)})
\prod_{m=N_0 +2}^{\infty} \Gamma_{-}(q^{-m + \hf}) \ |0 \rangle
.}
$$
Substituting $\Psi_D(z) = \G_{-}^{-1}(z) \G_{+}(z)$ we obtain
$$
\eqalign{
& Z_D(q, N, N_0) =  \langle 0| \ \prod_{n=1}^{\infty} \Gamma_{+}(q^{n- \hf})
\ \times \cr
& \ \ \ \ 
\prod_{m=1}^{N_0 + 1} \Gamma_{-}(q^{-m + \hf}) \G_{-}^{-1}(q^{-(N + \hf)})
\G_{+}(q^{-(N + \hf)}) 
\prod_{m=N_0 +2}^{\infty} \Gamma_{-}(q^{-m + \hf}) \ |0 \rangle
.}
$$
Commuting first the $\G_{+}(q^{-(N + \hf)})$ to the right gives a factor
$$
\prod_{m=N_0 +2}^{\infty} (1 - q^{m-N-1})^{-1} = 
{1 \over L(-g_s(N+ \hf) + (N_0 +1) g_s, q)}.
$$ 
The leftover $\G_{-}^{-1}(q^{-(N + \hf)})$ cancels then the $N$th $\G_{-}$
in the sequence. The remainder is then the McMahon function, up to an
extra factor making up for the missing $\G_{-}$ at the place $N$. This
extra factor is
$$
\prod_{n=1}^{\infty} (1 - q^{n+N}) = 
L(g_s(N+ \hf), q).
$$
Thus 
\eqn\gbrane{
Z_{D}(q, N, N_0) = M(q) \ {L(g_s(N+ \hf), q) \over L(-g_s(N+ \hf) + (N_0 +1) g_s, q)
}.
}
Comparing with the wavefunctions appearing in the 3D correlators we
obtain
\eqn\gbranecomp{
Z_{D}(q, N, N_0) = M(q) \ \Phi_{3D} (g_s(N + \hf), N_0 +1).
}
Since we have started with a fermionic insertion at $u= g_s(N+ 1/2)$ and
at the slice $t=N_0+1$, there is a precise agreement. The branes inserted in
the 3D correlator with wavefunction $\Phi_{3D}(u, t)$ correspond to the 
fermionic insertions $\Psi(e^{g_s u})$ at the slice $t$. 
Note that we have the choice to insert the fermion at a slice $t\ge 0$,
or at $t<0$. Inserting at $t<0$ precisely reproduces $\Phi_{3D}(u, t)$ 
for $t<0$. 
The branes are generalized branes, in the sense that they depend on
the two parameters $u$ and $t$.

It is interesting to note that the Fourier transform of the wavefunction
of the generalized branes $\Phi_{3D}(u, t)$ is the Wigner function.
Let us take $t>0$. Using that the generalized brane wavefunction 
is a product of a brane and a shifted antibrane wavefunction, the Fourier
transform gives
$$
W(g_s t, y) = \int d u L(u, q) L^{*} (-u + g_s t, q) e^{-i u y},  
$$
where we recognize the Wigner function. The Wigner function
appeared in the recent black hole interpretation of the topological
string partition function \bh. It would be interesting to investigate this
connection further.

\subsec{Universality for 3D partitions}

It is interesting to observe that a universal scaling behaviour also appears in
the case of the correlators of the 3D random partitions. In the same scaling
limit as in the two-dimensional case, the 3D kernel reduces to the incomplete
$\beta$-kernel \okounkov
$$
K_{\beta}(t, h) = {1 \over 2 \pi i} \int_{{\bar \eta}}^{\eta} d z (1 - z)^{t}
{1 \over z^{h + t/2 +1}}.
$$
where the distances kept fixed are $t=t_i -t_j$ and $h=h_i -h_j$, and 
$\eta$ is a density scaling parameter.
This reduces to the plane sine-kernel on two different slices. If we take
$t=0$, or $x=y$ in the original coordinates
$$K_{\beta}(h) = {\sin{h \phi} \over \pi h},$$
we obtain the sine-kernel in $h$. Taking $h+t/2 +1=0$, or $x-z =1$, 
and changing variables $z \rightarrow (1-z)$
$$ 
K_{\beta}(t, \tilde{\phi}) = {\sin{\tilde{\phi}(t+1)} \over \pi (t+1)},
$$
the kernel reduces to a sine kernel in $t'=t+1$. Here $\tilde{\phi}$ is
a redefined scaling variable. 
This means the kernel is likely connected to a two-matrix model. Taking
a Gaussian two-matrix model with 
$$ 
\int d X d Y \  e^{ -{1 \over g_s} (\hf X^2 + \hf Y^2  + V(X, Y))}, 
$$
the universal kernel has the form of $K(x_i -x_j, y_i-y_j) = K(x,y)$.
Reducing to $y=0$ or $x=0$ we have 
$$
\eqalign{
K(x,0) &\sim {\sin{\pi \rho x} \over  \pi x}, \cr
K(0,y) &\sim {\sin{ \pi \rho y} \over  \pi y}}.
$$
Thus reducing to the two slices would correspond to restricting to the
eigenvalues of only $X$ or only $Y$ in the two-matrix model. It is clear that
the corresponding two-matrix model must have an interaction term, otherwise
the full correlator would be just a product of the two single-matrix
(or single-slice) correlators. This is apparently not the case for the
$\b$-kernel. 

The two-matrix model in question is most likely the one related to the
Chern-Simons partition function \cs\ and the topological vertex \mcs.

To understand the full kernel, we need the four-point function
in this two-matrix model, and its scaling limit.
Such kernels has been considered in the context of Brownian motion \tw\ 
for the usual two-matrix models.
Using the context of Brownian motion \sebas\ could lead to a more 
precise identification.

The universal scaling formula (incomplete beta kernel) is a probe of the
crystal at short distances. Translated to topological
strings, it may indicate an interesting short distance scaling regime 
of string amplitudes.
 
\bigskip
\centerline{\bf Acknowledgements}

We would like to thank M.~Mari\~no, E.~Verlinde and C.~Vafa for
discussions. This research is supported by NWO, the FOM Programme {\it
String Theory and Quantum Gravity}, and the CMPA grant of the
University of Amsterdam.

\appendix{A}{Universal kernel from Hermite polynomials}

The universal kernel and its endpoint form can also be computed using
orthogonal polynomials 
$$ \int d \l \ e^{-{V(\l) \over g_s}} P_n (\l) P_m(\l) = \delta_{nm}. $$
Defining the wavefunctions
$$ \Psi_n(\l_i) = P_n(\l_i)\  e^{-{V(\l) \over 2 g_s}}, $$
and using the Darboux-Christoffel formula, the kernel (up to a constant) 
can be expressed as
\eqn\DC{ 
K(\l_i, \l_j) = {1 \over N} \sum_{n=0}^{N-1} 
\Psi_n(\l_i) \Psi_n(\l_j) 
=  { \Psi_N(\l_1) \Psi_{N-1}(\l_2) - \Psi_{N-1}(\l_1) 
\Psi_N(\l_2)  \over \l_1 - \l_2}. }

For the Gaussian potential we have
$$ \Psi_n(x) = {1 \over (2 g_s \pi)^{1 \over 4} 2^{n \over 2} \sqrt{n!}}
\ H_n \left({x \over \sqrt{2 g_s}} \right) e^{-{x^2 \over 4 g_s}} $$
where $H_n$ are the Hermite polynomials. These are the wavefunctions
of the harmonic oscillator with $\hbar = 2 g_s$. 
Using $H_n'(x) = 2 n H_{n-1}(x)$ we can rewrite \DC\ as
\eqn\Herm{\k(\l_i, \l_j) = {A_N A_{N-1} \over 2 N} e^{-{\l_1^2 
\over 4 g_s} - {\l_2^2 \over 4 g_s}} \left( {H_{N}({\l_1 \over \sqrt{2 g_s}})
H_{N}' ({\l_2 \over \sqrt{2 g_s}}) - H_{N} ({\l_2 \over \sqrt{2 g_s}})
H_{N}' ({\l_1 \over \sqrt{2 g_s}}) \over \l_1 - \l_2} \right) }
where $A_n$ is the normalization factor of the Gaussian wavefunctions
$$A_n = {1 \over (2 g_s \pi)^{1 \over 4} 2^{n \over 2} \sqrt{n!}}.$$

We need $\k(\l_i, \l_j)$ in the limit $N \rightarrow \infty$, $g_s \rightarrow
0$, $\mu = g_s N$ fixed and large. To find the asymptotic expansion of the 
Hermite polynomials, it is useful to rewrite them in terms of parabolic
cylinder functions $U(a,x)$ as
$$H_n(x) = 2^{n/2} \ e^{{1 \over 2} x^2} \ U(-n-{1 \over 2}, \sqrt{2} x).$$
The asymptotic expansion of the parabolic cylinder function $U(a,x)$ is 
given in terms of the quantity $Y = \sqrt{4 |a| - x^2}$. Since $a= -N-1/2$,
we expand in the region where $a$ is large, negative. In this region the
argument $\l_i / \sqrt{g_s} \sim \sqrt{N \over \mu} \l_i  \sim \sqrt{N}$,
so $x$ is moderately large. Using the asymptotic expansion in this regime,
and neglecting lower order terms in the $1 \over N$ expansion, we find
$$ U(-N-1/2,x) = {2 \sqrt{N!} \over (2 \pi)^{1/4}} {1 \over \sqrt{Y(x)}}
\ \cos{ \left({1 \over 2} \int_{0}^x Y(x') d x' - {N \pi \over 2} \right)}. $$
Substituting in \Herm, and expanding when $\l_i - \l_j \sim O(1/N)$ we
arrive at
$$K(\l_i, \l_j) \sim 
{{\sin{N \pi \rho(\bar{\l}}) (\l_i - \l_j)} \over \pi (\l_i - \l_j)},$$
where $\bar{\l} = (\l_i + \l_j)/ 2$, and $\rho(\bar{\l})$ is the 
normalized eigenvalue density
$$\rho(\bar{\l}) = {1 \over 2 \mu \pi} \sqrt{4 \mu - \bar{\l}}.$$
The overall normalization factor can be fixed from the definitions, using
that $\k(\l, \l) = \rho(\l)$. 
Finally $$\k(\l_i, \l_j) =  {{\sin{N \pi \rho(\bar{\l}}) (\l_i - \l_j)} \over
N \pi (\l_i - \l_j)},$$
in agreement with \uc.

\subsec{Endpoint asymptotics}

When scaling close to one of the branch points, $\l \rightarrow \pm
2 \sqrt{\mu}$, the kernel has a different asymptotic expansion in terms
of Airy functions. Let us scale to the positive branch point 
$2 \sqrt{\mu}$. At large $a$ the parabolic cylinder functions can be
expressed in terms of Airy functions as 
$$U(a, x) \sim 2^{-{1\over 4} - {1 \over 2} a} \G\left({1 \over 4} - 
{1 \over 2} a \right) \left({t \over \xi^2 -1}\right)^{1 \over 4} 
{\rm Ai}(t),$$
where
$$ x = 2 \sqrt{|a|} \xi, \quad t = (4 |a|)^{2/3} \t, $$
and
$$\t = \left( {3 \over 8} ( \xi \sqrt{\xi^2 -1} - \cosh^{-1}{\xi} )
\right)^{2\over3}.$$ 
We need the scaling limit when $x \rightarrow 2 \sqrt{|a|}$, that is
when $\xi \rightarrow 1$. In this limit
$$ U(a, x) \sim 2^{-{1\over 4} - {1 \over 2} a} \G\left({1 \over 4} - 
{1 \over 2} a \right)  |a|^{1 \over 6} \ \rm{Ai}(t),$$
where 
$t = 2 |a|^{2 \over 3} (\xi -1)$
mesures the distance from the branch point. Close to the branch point
then the Hermite polynomials have the asymptotics
$$
H_{N}(x) \sim  2^{N \over 2} e^{x^2 \over 2} 
\left(N + {1 \over 2}\right)^{1 \over 6} 
\G\left( {1 + N \over 2}\right) {\rm Ai}(u), 
$$
where
$$
u = \left(N + {1 \over 2} \right)^{1 \over 6} \left( x - \sqrt{4 N + 2} 
\right).
$$
Substituting in \Herm\ we find
$$K(u_1, u_2) \sim { {\rm Ai(u_1)} {\rm Ai'(u_2)} - 
{\rm Ai(u_2)} {\rm Ai'(u_1)} \over u_1 - u_2 }. $$
(Here we rescaled $u$ by a $1 / \sqrt{g_s}$ as doing the substitution.)

\listrefs
\end